# Mössbauer studies of the peculiar magnetism in parent compounds of the iron-based superconductors

A. K. Jasek[1], K. Komędera[1], A. Błachowski[1], K. Ruebenbauer[1*], J. Żukrowski[2,3], Z. Bukowski[4], and J. Karpinski[5,6]

[1]Mössbauer Spectroscopy Division, Institute of Physics, Pedagogical University
*ul. Podchorążych 2, PL-30-084 Kraków, Poland*

[2]AGH University of Science and Technology, Academic Center for Materials and Nanotechnology
*Av. A. Mickiewicza 30, PL-30-059 Kraków, Poland*

[3]AGH University of Science and Technology, Faculty of Physics and Applied Computer Science, Department of Solid State Physics
*Av. A. Mickiewicza 30, PL-30-059 Kraków, Poland*

[4]Institute of Low Temperature and Structure Research, Polish Academy of Sciences
*ul. Okólna 2, PL-50-422 Wrocław, Poland*

[5]Laboratory for Solid State Physics, ETH Zurich
*CH-8093 Zurich, Switzerland*

[6]Institute of Condensed Matter Physics, EPFL
*CH-1015 Lausanne, Switzerland*

[*]Corresponding author: sfrueben@cyf-kr.edu.pl

## Abstract

A review of the magnetism in the parent compounds of the iron-based superconductors is given based on the transmission Mössbauer spectroscopy of $^{57}$Fe and $^{151}$Eu. It was found that the 3d magnetism is of the itinerant character with varying admixture of the spin-polarized covalent bonds. For the '122' compounds a longitudinal spin density wave (SDW) develops. In the case of the EuFe$_2$As$_2$ a divalent europium orders anti-ferromagnetically at much lower temperature as compared to the onset of SDW. These two magnetic systems remain almost uncoupled one to another. For the non-stoichiometric Fe$_{1+x}$Te parent of the '11' family one has a transversal SDW and magnetic order of the interstitial iron with relatively high and localized magnetic moments. These two systems are strongly coupled one to another. For the "grand parent" of the iron-based superconductors FeAs one observes two mutually orthogonal phase-related transversal SDW on the iron sites. There are two sets of such spin arrangements due to two crystallographic iron sites. The FeAs exhibits the highest covalency among compounds studied, but it has still a metallic character.

## 1. Introduction

This contribution is a review of the results obtained up to now by means of the transmission Mössbauer spectroscopy as far as magnetic order is considered in the parent compounds of the iron-based superconductors [1-4]. Majority of the iron-based superconductors is characterized by presence of the iron-pnictogen or iron-chalcogen layers with tetrahedrally coordinated iron by covalent partner [5]. Highly ordered structures appear for the '122' family with two inverted one versus another iron bearing layers in the chemical cell [6]. Atoms responsible for spacing of above layers can belong to the rare earths bearing their own localized magnetic moments e.g. europium [7]. On the other hand, the 3d magnetism induced by iron has highly itinerant character [8]. Compounds of the '11' family form simpler structures with one iron bearing layer per cell, but they have many vacant interstitial positions susceptible to the occupation by iron [9]. Finally, one has to look at the "grand parents" of above compounds e.g. for the FeAs with the octahedral iron coordination and bonds being intermediate between metallic and covalent [10].

The paper is organized as follows. Section 2.1 deals with the longitudinal spin density waves (SDW) in the $EuFe_2As_2$ parent compound of the '122' iron-based superconductor family. Magnetic order of $Eu^{2+}$ is discussed as well. A transversal SDW of $Fe_{1+x}Te$ is discussed in section 2.2 together with the magnetism of interstitial iron occupying vacant positions. The last section 2.3 is devoted to the peculiar magnetic order of FeAs. There are two mutually orthogonal transversal SDW on two distinctly different iron sites leading to the complex magnetic spirals incommensurate with the corresponding lattice period. These spirals are seen as ellipses by the neutron scattering method [10], while they reveal more complex structure observed via the hyperfine field – namely the fourth order deformation [4].

## 2. Review of the results

### 2.1. Magnetism of $EuFe_2As_2$

This compound crystallizes in the tetragonal structure with the Fe-As layers being perpendicular to the longest *c*-axis. Magnetic moments of the 3d electrons order at 192 K with the simultaneous slight distortion of the crystal symmetry to the orthorhombic one. A longitudinal SDW propagates along the *a*-axis and the order has anti-ferromagnetic character. The SDW propagation vector is incommensurate with the inverse lattice period along the *a*-axis. Figure 1 shows selected $^{57}Fe$ (14.41-keV resonant transition) transmission Mössbauer spectra versus temperature. The shape of SDW along the propagation direction is shown for one period in Figure 2 for selected temperatures together with the amplitudes of the first two dominant harmonics and maximum value of the hyperfine field. A development of the magnetic order starts above transition temperature in the narrow incoherent region. A SDW develops as magnetically ordered "sheets" parallel to the *b-c* plane. They have roughly triangular shape along the propagation direction, i.e. *a*-axis. They fill the whole available space upon lowering of the temperature and gradually transform into rectangular form. Hence, they are almost indistinguishable from the simple anti-ferromagnetic ordering close to the ground state as long as the Mössbauer spectroscopy is concerned [1].

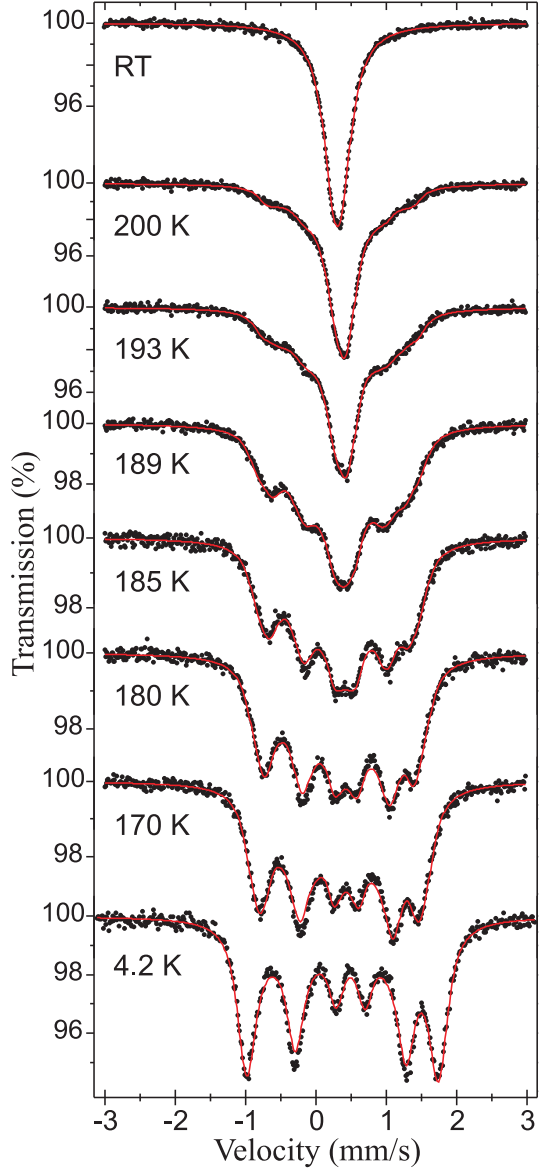
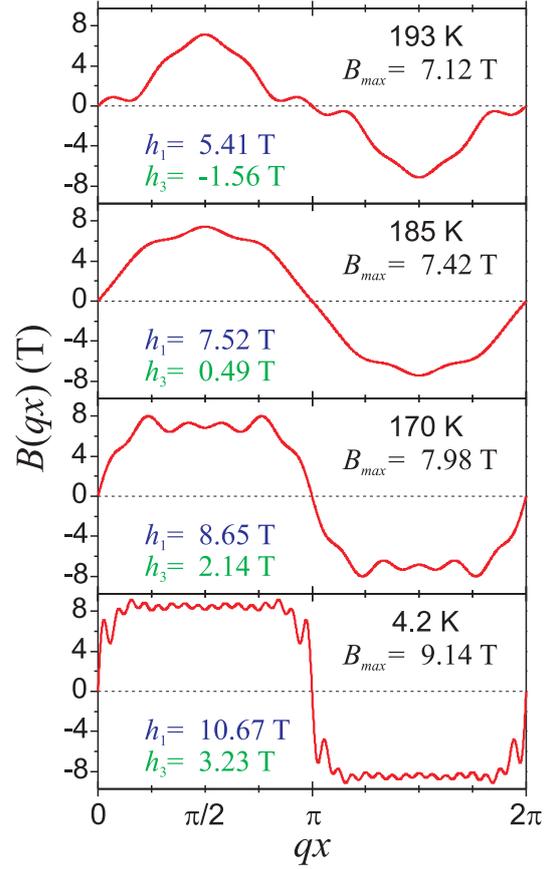

**Figure 2** SDW shapes along the normalized $qx$ propagation direction for the EuFe$_2$As$_2$. Maximum hyperfine field $B_{max}$ and amplitudes of the first two dominant harmonics $h_1$ and $h_3$ are shown.

**Figure 1** $^{57}$Fe Mössbauer spectra of the EuFe$_2$As$_2$.

Europium atoms occur in the divalent state, and hence they have formally $^8S_{7/2}$ electronic configuration with the large magnetic moment due to the spin configuration and almost negligible orbital moment. They remain in the disordered magnetic state all the way down to 19 K and order in the anti-ferromagnetic fashion below with magnetic moments being parallel to the *a*-axis. Selected $^{151}$Eu transmission spectra (21.6-keV resonant transition) are shown in Figure 3. Hence, one can conclude that the SDW very weakly interact with the localized europium moments, i.e., the SDW sheets described above have "holes" around europium atoms [2].

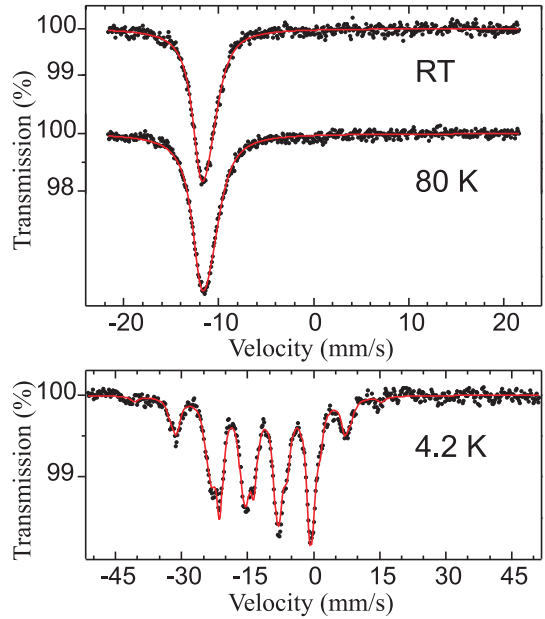

**Figure 3** $^{151}$Eu Mössbauer spectra of the EuFe$_2$As$_2$. Note europium hyperfine field of 27.4 T at 4.2 K.

*2.2. Magnetism of Fe$_{1+x}$Te*

Compounds of the '11' family cannot be made stoichiometric, as iron bearing layers are coupled one to another by some interstitial atoms, the latter intercalated between subsequent layers. Hence, one can see some ordered vacancy lattice filled more or less randomly by intercalated atoms. A good example of such structure is Fe$_{1+x}$Te with excess iron being intercalated between regular layers described above. A magnetic order sets about 70 K for the low non-stoichiometry parameter x and ordering temperature lowers with the increasing x till x=0.14 recovering partially for still higher x [9, 11]. Similar SDW magnetic order occurs as for the '122' compounds, but SDW has transversal character in the present case. Interstitial iron is very similar to the high spin trivalent iron. It couples quite strongly to SDW and therefore orders magnetically at the onset of SDW. The hyperfine field of intercalated iron depends on the occupation of adjacent vacancies by other (intercalated) iron atoms. The highest field is observed for "isolated" iron atoms. Figure 4 shows $^{57}$Fe transmission Mössbauer spectra at 65 K (just below magnetic transition temperature) and 4.2 K for selected departure from stoichiometry x. Corresponding shape of SDW along the propagation direction is shown as well together with the amplitudes of two dominant harmonics and maximum hyperfine field. Contributions due to the regular iron and three distinct sites of iron in the interstitial positions are shown, too. Note the large hyperfine fields for the interstitial iron – especially for the "isolated" iron atom [3]. The influence of the interstitial iron on the SDW shape changes between x=0.10 and x=0.14 in accordance with the neutron diffraction data [9]. In general, the SDW becomes more irregular with the increasing concentration of the interstitial iron. The increase of above concentration leads to the stronger coupling between regular Fe-Te layers observed as the shrinking of the cell in the *c*-axis direction [9]. For large concentration of the interstitial iron one cannot obtain superconductivity as these disordered atoms bearing quite large and localized magnetic moments destroy Cooper pairs. Hence, one has to remove excess of the interstitial iron by substitution of Te by either S or Se [12] and de-intercalation e.g. applying mild organic acids being components of the alcoholic beverages [13, 14].

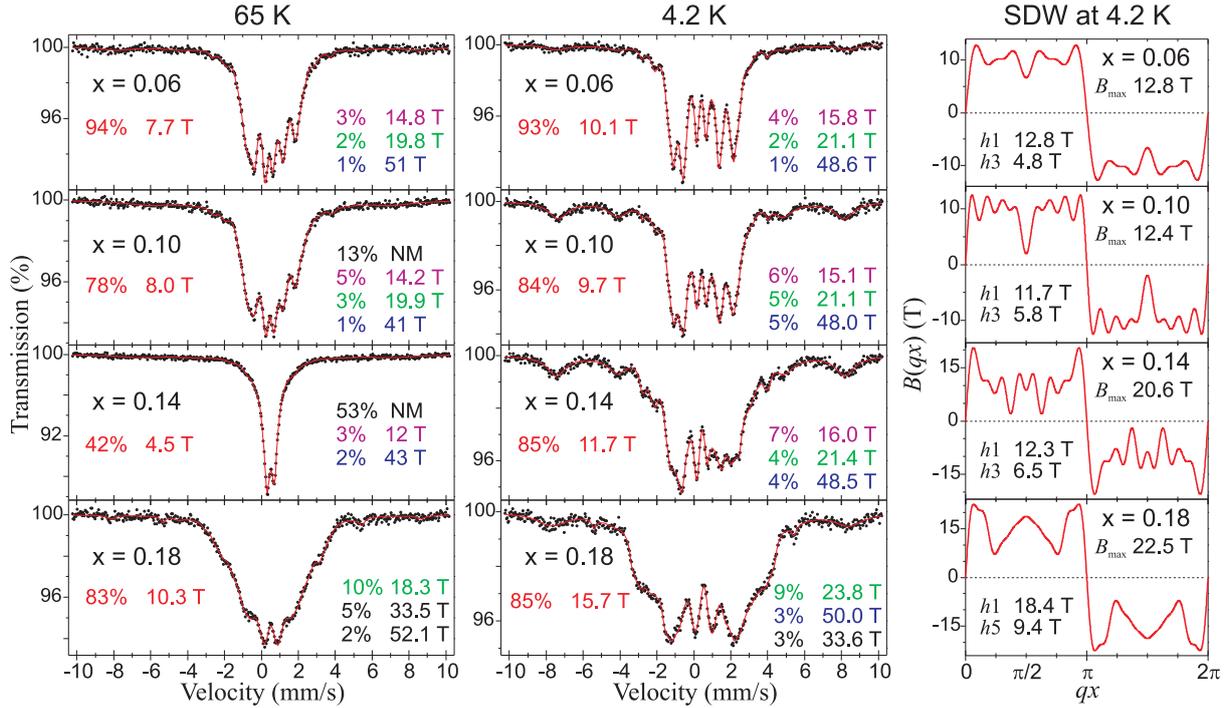

**Figure 4** $^{57}$Fe Mössbauer spectra of the Fe$_{1+x}$Te at 65 K and 4.2 K for various values of the non-stoichiometry parameter x. Contributions due to the regular iron and mean squared hyperfine field of SDW are shown at left. Contributions due to three types of the interstitial iron atoms with the corresponding hyperfine fields are shown at right with the symbol NM (at 65 K) denoting non-magnetic component. For x=0.18 one has some additional components due to unreacted α-Fe (~33 T) and spurious Fe$_3$O$_4$ (~52 T). The right column shows SDW shape at 4.2 K with the maximum field $B_{max}$ and amplitudes of the first two dominant harmonics.

*2.3. Spin spirals in FeAs*

An iron mono-arsenide FeAs crystallizes in the orthorhombic structure as shown in Figure 5 [4]. Basically, one can interpret diffraction data within the *Pnma* symmetry group [15]. All four atoms of iron within the chemical unit cell are equivalent within the *Pnma* group. Each iron atom is coordinated by six arsenic atoms and within the *Pnma* group there are four different distances between iron and adjacent arsenic as two pairs of bonds lead to the same distances within the pair. However, a distortion leading to the lower symmetry *Pna2$_1$* has been reported [15]. A distortion has probably more or less random character and leads to the iron displacements shown by arrows of Figure 5. Distortions shown in "black" and "gray" are equally probable. For such case one has two pairs of iron atoms per chemical cell (shown in "red" and "blue" in Figure 5) mutually inequivalent, i.e. "red" iron 1 located near $[0\ k+\tfrac{1}{2}\ 0]$ planes and "blue" iron 2 at $[0\ k\ 0]$ planes with $k$ denoting the Miller index. All iron-arsenic nearest neighbor distances become different one from another under such circumstances – for both iron sites. Note that the mirror symmetry shown as heavy rectangle in Figure 5 is broken under above distortion. High temperature Mössbauer spectra confirm presence of above distortion, as they are composed of two equal intensity doublets differing by the isomer shift one from another [4]. A difference is seen up to about room temperature. The system remains metallic at all temperatures and there are no structural phase transitions either versus temperature or versus pressure. A magnetic moment per iron atom is close to the magnetic moment of the high spin trivalent iron at temperatures well above room temperature and it lowers with decreasing temperature indicating increasing metallicity at low temperature [16]. One can destroy iron magnetic moments applying pressure of about 11 GPa [17]. The system orders anti-ferromagnetically at about 70 K, i.e. before having lost magnetic moment entirely.

The magnetic order stops further reduction of the iron magnetic moment with lowered temperature. Magnetic moments order in the *a-b* plane. A transition is extremely sharp despite lack of the structural changes. Hence, neither the magnetic moment nor the hyperfine magnetic field is correct order parameter. The iron dynamics exhibits *quasi*-harmonicity above room temperature and it is harmonic at lower temperature with the iron motion being more restricted along the *b*-axis than in other directions. Iron lattice hardens upon transition to the magnetically ordered state and the iron motion becomes isotropic. Iron dynamics is the same for all iron sites within the chemical cell. Selected Mössbauer spectra versus temperature are shown in Figure 6. A magnetic order leads to the development of the SDW propagating along the *c*-axis. They are incommensurate with the respective lattice period. One of them propagates through iron 1 ("red" - $[0\,k+\frac{1}{2}\,0]$), while the other one through iron 2 ("blue" - $[0\,k\,0]$). Each of them is transversal and composed of two planar SDW with respective fields pointing along the *a*-axis and *b*-axis, respectively. Hence, one obtains a complex spiral pattern [18, 4] shown in Figure 7 for selected temperatures and for both iron sites. The „blue" pattern is sharper as it is not perturbed by the lattice distortion lowering the symmetry from the *Pnma* to *Pna2$_1$* group. Both patterns require presence of the quartic terms indicating significant contribution of the covalent bonds to the 3d magnetism (squared wave functions of the d state). It is likely that these covalent effects protect the system from the complete loss of the magnetic moment upon lowering temperature. It seems that magnetic order leads to the larger coherence length as far as distortions lowering the *Pnma* symmetry are considered. It is worth noticing that the magnetic moment along the *b*-axis was found to be 15 % larger than the magnetic moment along the *a*-axis by using the polarized neutron scattering method [10].

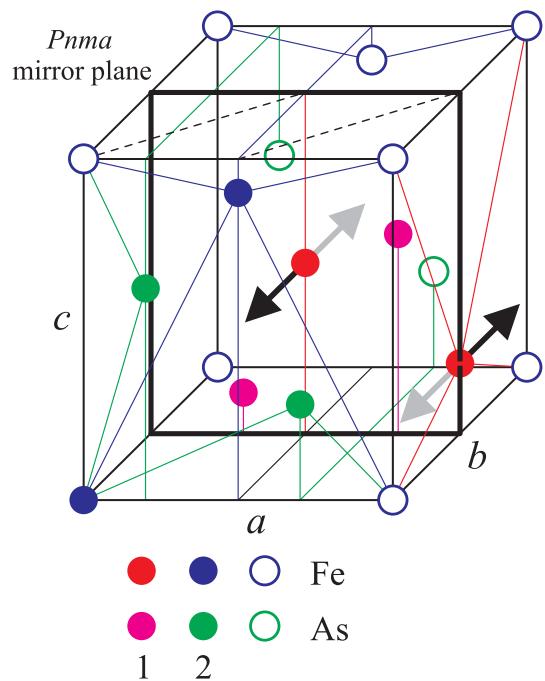

**Figure 5** Chemical unit cell of the FeAs. The longest lattice constant is along the *c*-axis, while the shortest is along the *b*-axis. Black and gray arrows show possible departures from the *Pnma* symmetry. The mirror plane exists for the *Pnma* symmetry. Empty circles denote atoms belonging to the adjacent cells. Red and blue iron atoms are equivalent without symmetry breaking distortion.

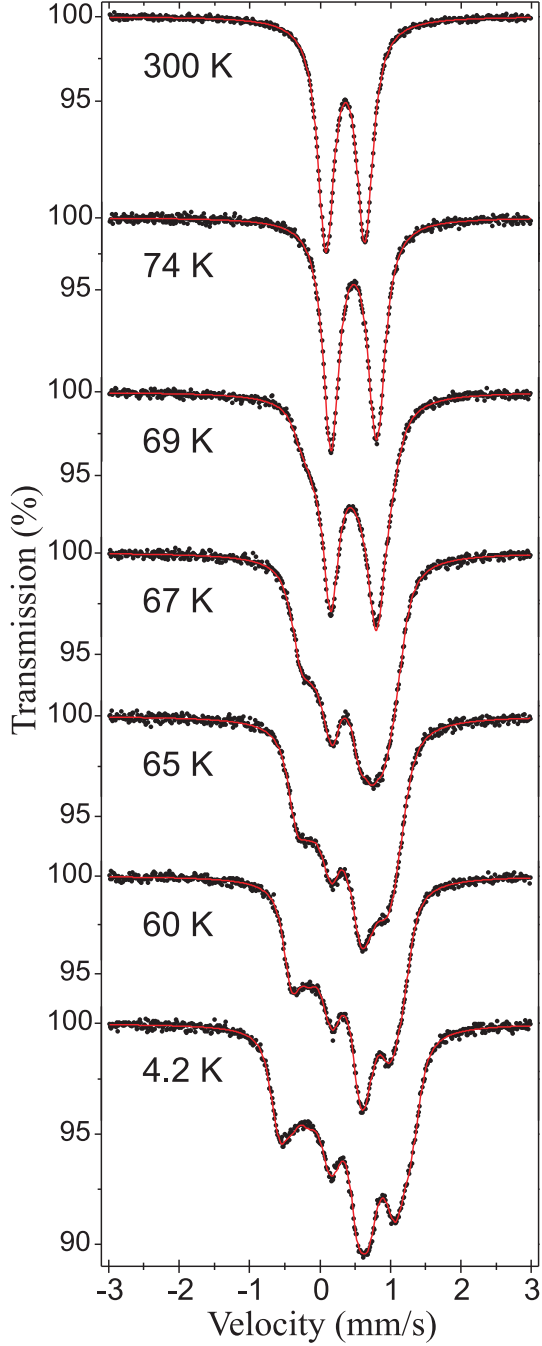

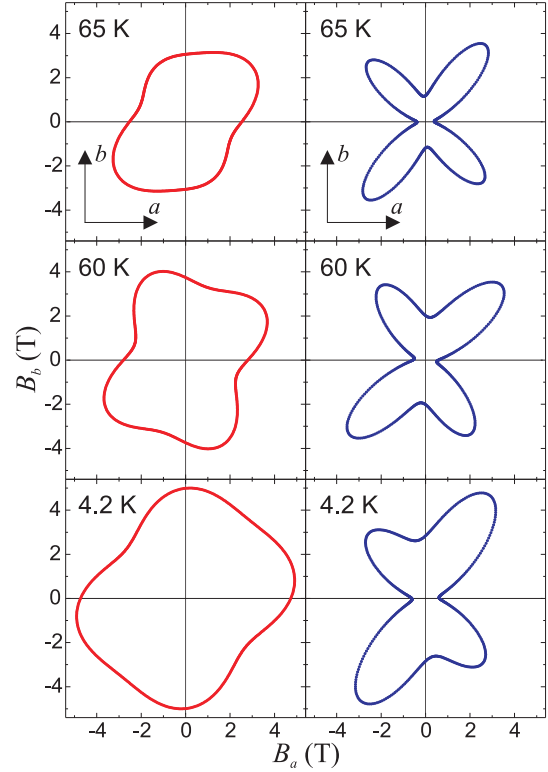

**Figure 7** Spirals made by the iron hyperfine magnetic field on the *a-b* plane for the FeAs. Red spirals are due to $[0\ k+\frac{1}{2}\ 0]$ iron, while the blue spirals are due to $[0\ k\ 0]$ iron. The symmetry of the blue spirals closely follows directions of the Fe-As bonds projected on the *a-b* plane i.e. the field has maxima almost along the projected bonds.

**Figure 6** $^{57}$Fe Mössbauer spectra of the FeAs.

## 3. Conclusions

Magnetism of the 3d electrons in the parent compounds of the iron-based superconductors has definitely itinerant character with the admixture of the spin-polarized covalent bonds. Hence, one obtains complex shape of SDW. Rare earth moments remain localized and they are weakly coupled to SDW. For *quasi* two-dimensional systems like the '122' compounds the SDW are longitudinal. Increased coupling between iron bearing layers transforms SDW into transversal. For the complete loss of the low dimensionality one obtains two mutually orthogonal and transversal SDW with the definite phase relationship between orthogonal components. Therefore complex spirals are observed. Increased admixture of covalent bonds introduces quartic terms to the SDW patterns and differentiates SDW patterns for various crystallographic sites. Spin-polarized covalent bonds (s, and anisotropic p and d electrons) lead to the strong magneto-elastic effects. Iron intercalated between iron bearing layers approaches the high spin trivalent iron provided it is almost isolated in the vacancy lattice.

In order to achieve superconductivity one has to destroy 3d magnetic moments applying either doping (n-type, p-type or isovalent) or external pressure. It is important to remove majority of the localized 3d moments. Localized 4f moments weakly couple to the Cooper pairs, and therefore a magnetic order of the 4f kind moments could coexist with the superconducting state within the same electronic system [2, 19]. However, no system with the 3d magnetic order and superconductivity within the same electronic system has been found by the Mössbauer spectroscopy up to now [20-23].

**Acknowledgments**

Dr. Paweł Zajdel, Division of Physics of Crystals, Institute of Physics, Silesian University, Katowice, Poland is thanked for preparation of the $Fe_{1+x}Te$ samples.
This work was supported by the National Science Center of Poland, Grant DEC-2011/03/B/ST3/00446. J. K. acknowledges support by the SNSF (Project No. 140760) and FP7 Super-Iron Project.